%% file: main.tex
\documentclass[
  journal=jcisd8, 
  manuscript=article
]{achemso}

\usepackage{amsmath,amssymb,bm}
\usepackage{graphicx,booktabs,array}
\usepackage[hidelinks]{hyperref}

\newcommand{\momentunit}{\mu_{\mathrm B}/\mathrm{f.u.}}

\title{Descriptor Non-Uniqueness in Symbolic Regression: Conditional Bayesian Analysis of SISSO Supports}

\author{Satadeep Bhattacharjee}
\affiliation{Indo Korea Science and Technology Center, Bangalore, India}
\email{s.bhattacharjee@ikst.res.in}

\begin{document}

\begin{abstract}
Compact analytical descriptors connect materials properties with interpretable combinations of physical and chemical variables, yet different expressions can capture similar trends in small datasets. We present a conditional Bayesian framework for quantifying competition among symbolic descriptors retained by the Sure Independence Screening and Sparsifying Operator (SISSO) method. Exact enumeration of fixed-size supports within the screened feature union yields posterior support probabilities, feature inclusion probabilities, and Bayesian-model-averaged predictions. Applied to magnetic moments of 88 full-Heusler alloys, the framework reveals substantial descriptor non-uniqueness even when the maximum-a-posteriori support coincides with deterministic SISSO. At both two and three descriptor dimensions, eight supports are required to accumulate 95\% posterior probability. Two prominent two-dimensional coordinates exhibit a Pearson correlation of 0.943 despite their distinct algebraic forms. Five-fold validation with screening repeated within each training fold shows higher prediction errors for Bayesian model averaging and undercoverage of nominal predictive intervals, separating descriptor competition from predictive performance. By making statistically competitive expressions explicit, the framework complements deterministic symbolic regression and provides a probabilistic basis for interpreting composition–property relationships in materials modeling.
\end{abstract}

\maketitle
\section{Introduction}
Symbolic regression seeks compact analytical relationships between materials properties and a small number of input variables. Such relationships are attractive when available datasets are limited and the objective includes interpretation as well as prediction. The Sure Independence Screening and Sparsifying Operator (SISSO) method makes this search feasible across extremely large libraries of nonlinear expressions by combining screening with sparse optimization.\cite{ouyang2018sisso} Applications to Heusler magnetic moments, electrocatalysis, superconductivity, and topological materials illustrate the range of problems accessible to this approach.\cite{bhattacharjee2022general,ram2023combining,xie2019functional,cao2020artificial}

A compact expression is not necessarily a uniquely identified relationship. Elemental properties such as size, valence, and electronegativity are correlated, and symbolic combinations can preserve or amplify these dependencies. On a small dataset, algebraically different expressions can therefore organize compounds similarly and yield comparable fits. Selecting one support by minimum training error hides this competition. Distinguishing a stable chemical trend from an unstable algebraic representation is particularly important when a descriptor is subsequently used to motivate a microscopic mechanism or to prioritize unfamiliar compounds.

Several strategies address uncertainty in symbolic models at different stages of the workflow.\cite{jin2019bayesianSR,pmlr-v337-boussif26a} Minimum-description-length selection and Bayesian symbolic regression compare alternative functional forms, including expression-tree representations.\cite{desmond2026exhaustive,10.1098/rsta.2025.0089} For SISSO, Nair et al. combine bootstrap ensembles with primary-feature dropout, allowing the selected symbolic representation to vary across fits.\cite{nair2025materials} Their SARBO framework instead uses adaptive SISSO representations within Bayesian optimization.\cite{nair2026interpretable} These approaches concern broader representation searches or repeated fitting. A complementary question is how probability is distributed among supports already retained by one residual-driven SISSO calculation.

Bayesian variable selection provides a statistical setting for examining competition among candidate supports: each support is assigned a posterior probability conditional on a likelihood and a prior.\cite{mitchell1988bayesian,george1997approaches} A maximum-a-posteriori (MAP) support is the single support with the greatest posterior probability. In contrast, the posterior inclusion probability (PIP) of a feature sums the probabilities of all supports that contain it and therefore describes marginal inclusion across the candidate models.\cite{mitchell1988bayesian,barbieri2004optimal} These summaries answer different questions; the highest-probability model need not be the best single model for prediction under a specified loss function.\cite{barbieri2004optimal}

Bayesian model averaging (BMA) instead combines predictions across models using their posterior probabilities, accounting for model uncertainty within the specified model space.\cite{raftery1997bma,hoeting1999bayesian,clyde2004model} The resulting probabilities depend on the assumed prior and the representation of the model space. Related predictors can be incorporated explicitly in support priors,\cite{chipman1996related} and model-space prior choices affect selection and multiplicity adjustment.\cite{scott2010multiplicity} These broader results motivate careful interpretation of PIPs when many correlated symbolic expressions are available.

The present analysis addresses the conditional question after SISSO screening using established conjugate Gaussian subset inference. Standard SISSO constructs the candidate union without modification; all supports of a prescribed dimension are then enumerated within that union. MAP, PIPs, and BMA summarize support competition and its predictive consequences within this retained space. Exactness refers to coefficient integration and support enumeration under the specified model. It does not extend to uncertainty in discarded expressions, screening, preprocessing, or estimated variance parameters.

Magnetic moments of 88 $\mathrm{X}_2\mathrm{YZ}$ full-Heusler alloys provide a case study with correlated elemental inputs and distinct Co-, Fe-, and Ni-based target distributions. The central result is that substantial posterior mass remains distributed among several supports even when the most probable support matches deterministic SISSO. A fold-local validation also exposes higher BMA prediction errors, interval undercoverage, and deterioration on increasing descriptor dimension. Together, these observations delimit the role of conditional support analysis: it provides an explicit account of representation ambiguity within the retained library, while leaving screening stability and predictive calibration as separate questions.
\section{Methods: Conditional Bayesian Support Analysis}\label{sec:methods}

\begin{figure*}[t]
\centering
\includegraphics[width=\textwidth]{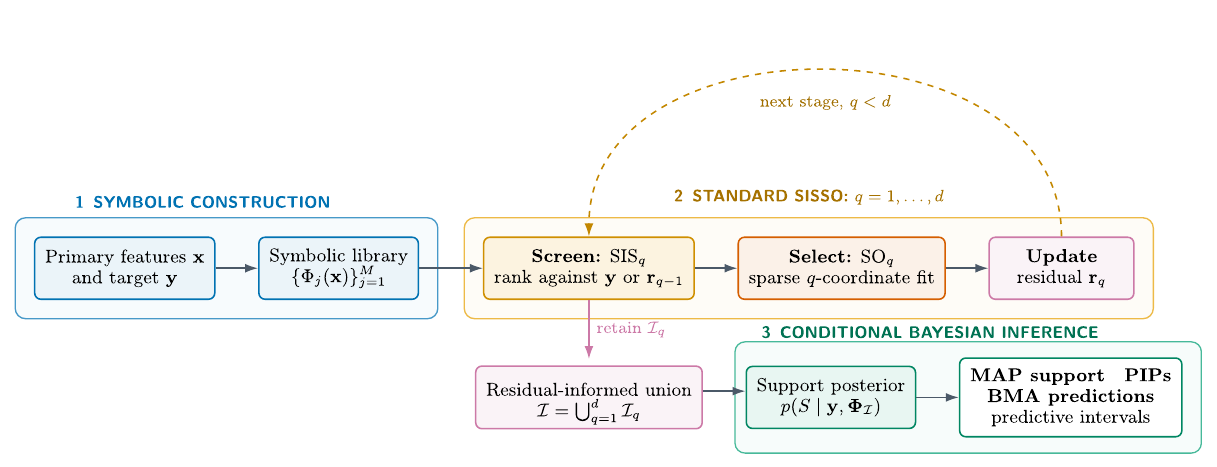}

\caption{Workflow of the support-uncertainty analysis. Standard SISSO retains its iterative target- and residual-driven sequence: each SIS step is followed by deterministic SO, and the resulting residual drives the next screen. The accumulated residual-informed union contains two SIS stages for $d=2$ and three for $d=3$. Only after this union has been constructed is Bayesian support inference applied. MAP denotes maximum a posteriori, PIP denotes posterior inclusion probability, and BMA denotes Bayesian model averaging.}
\label{fig:workflow}
\end{figure*}

\subsection{Standard SISSO and residual-driven screening}

The complete analysis sequence is summarized in Figure~\ref{fig:workflow}. Following Ouyang \textit{et al.}~\cite{ouyang2018sisso}, SISSO starts from a set of primary features and recursively applies algebraic and functional operators to generate a very large candidate feature space for symbolic regression. For a dataset of \(n_{\rm s}\) samples with targets \(y_i\), the resulting symbolic library can be written as
\begin{equation}
\{\Phi_j(\mathbf{x})\}_{j=1}^{M},
\end{equation}
where \(M\) may be enormous. In a regression setting, the target vector is approximated as
\begin{equation}
\mathbf{y}\approx \bm{\Phi}\mathbf{c},
\end{equation}
with sensing matrix \(\bm{\Phi}\in\mathbb{R}^{n_{\rm s}\times M}\) and sparse coefficient vector \(\mathbf{c}\). Direct optimization over all \(M\) candidates is generally intractable, so standard SISSO alternates screening and sparse fitting~\cite{ouyang2018sisso}. The first screen, \(\mathrm{SIS}_1(\mathbf y)\), ranks features against the target; \(\mathrm{SO}_1\) then selects the first coordinate and defines a residual \(\mathbf r_1\). The next screen, \(\mathrm{SIS}_2(\mathbf r_1)\), is therefore informed by what the first coordinate did not explain. This SIS--SO--residual sequence continues until dimension \(d\). Throughout this manuscript, \emph{standard SISSO} denotes this complete deterministic iterative procedure, whereas \emph{deterministic SO} denotes the sparse search at a particular stage.

For a selected stage-$q$ support $\gamma_q$, the fitted residual is
\begin{equation}
\mathbf r_q=\mathbf y-\widehat c_{0,q}\mathbf 1-\bm{\Phi}_{\gamma_q}\widehat{\mathbf c}_q,
\qquad
(\widehat c_{0,q},\widehat{\mathbf c}_q)
=\arg\min_{c_0,\mathbf c}\|\mathbf y-c_0\mathbf 1-\bm{\Phi}_{\gamma_q}\mathbf c\|^2.
\end{equation}
The score $s_j^{(q)}$ below denotes the relevance ranking returned by the SISSO screen; it is distinct from the feature-scaling factors $a_j$ used in Bayesian inference.

\subsection{Construction of the SISSO-screened support space}

Let \(s_j^{(q)}\) denote the SIS relevance score for feature \(\Phi_j\) at descriptor stage \(q\), evaluated against \(\mathbf y\) for \(q=1\) and the preceding residual for \(q>1\). The stage-specific screened index set is
\begin{equation}
\mathcal I_q=\operatorname{TopK}\left(\{s_j^{(q)}\}_{j=1}^{M};K_q\right).
\end{equation}
Bayesian analysis uses the accumulated union
\begin{equation}
\mathcal I=\bigcup_{q=1}^{d}\mathcal I_q,
\qquad
\bm{\Phi}_{\mathcal I}\in\mathbb{R}^{n_{\rm s}\times K},
\quad K=|\mathcal I|\ll M.
\end{equation}
For the full-data calculations, the stage contributions contain no duplicate expressions, giving $K=400$ from two 200-feature stages at $d=2$ and $K=300$ from three 100-feature stages at $d=3$. This disjointness is an observed property of these runs, not an assumption of the method. No probabilistic model is assigned to discarded features. Bayesian conclusions are therefore conditional on the residual-informed space constructed by standard SISSO.

\subsection{Bayesian support inference}

Bayesian inference is performed only after standard SISSO has constructed the accumulated screened space. For a candidate support
\begin{equation}
\gamma\subset\{1,\dots,K\},\qquad |\gamma|=d,
\end{equation}
we use the Gaussian linear model
\begin{equation}
\widetilde{\mathbf y}=\widetilde{\bm{\Phi}}_{\gamma}
\boldsymbol{\beta}_{\gamma}+\boldsymbol{\varepsilon},
\qquad
\boldsymbol{\varepsilon}\sim \mathcal N(\mathbf 0,\sigma^2\mathbf I),
\end{equation}
where the centered target and standardized design matrix are defined below. The number of rows in each inference calculation is denoted by $n$ ($n=88$ for a full-data fit and $n=n_{\rm train}$ within a validation fold). Active coefficients are assigned the prior
\begin{equation}
\boldsymbol{\beta}_{\gamma}\sim\mathcal N(\mathbf 0,\tau^2\mathbf I).
\end{equation}
Let $\widetilde{\mathbf y}$ be the training target centered by its training mean, and let $\widetilde{\bm{\Phi}}_{\gamma}$ contain the active symbolic columns centered and standardized by their training means $\overline{\boldsymbol\phi}_{\rm train}$ and standard deviations $\mathbf a_{\rm train}$. Near-constant columns are excluded using a relative numerical tolerance. The same training transformations are applied unchanged to validation samples, and the target intercept is restored after prediction. Standard SISSO itself is not standardized or otherwise altered. We write
\begin{equation}
\widetilde{\mathcal D}_{\mathcal I}=\{\widetilde{\bm{\Phi}}_{\mathcal I},\widetilde{\mathbf y}\},
\qquad
\mathcal C_{\mathcal I}=\{\mathcal I,\overline{\boldsymbol\phi}_{\rm train},
\mathbf a_{\rm train},\bar y_{\rm train},\sigma^2,\tau^2\},
\end{equation}
and condition every evidence, posterior, and predictive quantity below on $\mathcal C_{\mathcal I}$. Bayes' theorem then gives
\begin{equation}
p(\gamma,\boldsymbol{\beta}_{\gamma}\mid\widetilde{\mathcal D}_{\mathcal I},\mathcal C_{\mathcal I})
\propto
p(\widetilde{\mathbf y}\mid\gamma,\boldsymbol{\beta}_{\gamma},\mathcal C_{\mathcal I})
p(\boldsymbol{\beta}_{\gamma}\mid\tau^2)p(\gamma\mid\mathcal C_{\mathcal I}),
\end{equation}
where we define
\begin{equation}
\bm{\Lambda}_\gamma=
\frac{\widetilde{\bm{\Phi}}_\gamma^{T}
\widetilde{\bm{\Phi}}_\gamma}{\sigma^2}
+\frac{\mathbf I_d}{\tau^2},
\qquad
\mathbf h_\gamma=
\frac{\widetilde{\bm{\Phi}}_\gamma^{T}\widetilde{\mathbf y}}{\sigma^2}.
\end{equation}
The implementation evaluates the Gaussian integral in closed form as
\begin{align}
\log p(\widetilde{\mathbf y}\mid\gamma,\mathcal C_{\mathcal I})
={}-\frac12\Big[&n\log(2\pi)+n\log\sigma^2
+d\log\tau^2\nonumber\\
&+\log|\bm{\Lambda}_\gamma|
+\frac{\widetilde{\mathbf y}^{T}\widetilde{\mathbf y}}{\sigma^2}\nonumber\\
&-\mathbf h_\gamma^{T}\bm{\Lambda}_\gamma^{-1}\mathbf h_\gamma\Big].
\end{align}
The posterior support probability is therefore~\cite{george1993variable,george1997approaches},
\begin{equation}
p(\gamma\mid\widetilde{\mathcal D}_{\mathcal I},\mathcal C_{\mathcal I})=
\frac{p(\widetilde{\mathbf y}\mid\gamma,\mathcal C_{\mathcal I})
p(\gamma\mid\mathcal C_{\mathcal I})}
{\sum_{\gamma'}p(\widetilde{\mathbf y}\mid\gamma',\mathcal C_{\mathcal I})
p(\gamma'\mid\mathcal C_{\mathcal I})}.
\end{equation}

This is a support posterior conditional on the SISSO-screened union, the training preprocessing, the Gaussian coefficient prior, and the plug-in variance scales. It is not a posterior over the original unscreened symbolic library. The same training responses determine the screen and the conditional likelihood; no selective-inference adjustment is made for that selection.

For a validation fold, \(\sigma^2_{\rm fold}\) is the squared training RMSE of that fold's deterministic standard-SISSO model, and \(\tau^2_{\rm fold}=\operatorname{var}(\mathbf y_{\rm train})/d\), using the population variance. These quantities, the screened union, feature correlations, residuals, and all transformations use training information only. Full-data posterior diagnostics use the analogous full-data quantities and are not validation estimates.

\subsection{Posterior quantities: MAP, PIP, and BMA}

The MAP estimator is the support with the largest posterior probability; a PIP instead sums posterior weight over supports containing the feature.\cite{barbieri2004optimal,mitchell1988bayesian} The PIP of feature $k$ is
\begin{equation}
P(k\in\gamma\mid\widetilde{\mathcal D}_{\mathcal I},\mathcal C_{\mathcal I})
=\sum_{\gamma\ni k}p(\gamma\mid\widetilde{\mathcal D}_{\mathcal I},\mathcal C_{\mathcal I}),
\end{equation}
and predictions are obtained by Bayesian model averaging,\cite{raftery1997bma,hoeting1999bayesian}
\begin{equation}
p(y_*\mid\mathbf x_*,\widetilde{\mathcal D}_{\mathcal I},\mathcal C_{\mathcal I})
=\sum_{\gamma}p(\gamma\mid\widetilde{\mathcal D}_{\mathcal I},\mathcal C_{\mathcal I})
p(y_*\mid\mathbf x_*,\gamma,\widetilde{\mathcal D}_{\mathcal I},\mathcal C_{\mathcal I}).
\end{equation}
The MAP support need not equal deterministic SO. Deterministic SO minimizes its
$L_0$ search criterion, whereas the marginal likelihood contains goodness-of-fit
and model-volume contributions under a specified coefficient prior. In the
present low-dimensional spaces, all supports are enumerated. Predictive moments
are accumulated over every support with normalized probability at least
$10^{-15}$ rather than only the displayed top-support list; the omitted mass is
recorded for every calculation and retained weights are renormalized. Even the
worst-case omitted-mass bound at $d=3$ is below $4.46\times10^{-9}$.
Reported 95\% predictive intervals use a normal approximation constructed from
the exact first two moments of this support mixture; they are not exact mixture
quantiles.

Writing $w_\gamma=p(\gamma\mid\widetilde{\mathcal D}_{\mathcal I},\mathcal C_{\mathcal I})$, the coefficient posterior is
\begin{equation}
\boldsymbol\beta_\gamma\mid\widetilde{\mathcal D}_{\mathcal I},\mathcal C_{\mathcal I}
\sim\mathcal N(\mathbf m_\gamma,\mathbf S_\gamma),
\qquad \mathbf m_\gamma=\bm{\Lambda}_\gamma^{-1}\mathbf h_\gamma,
\quad \mathbf S_\gamma=\bm{\Lambda}_\gamma^{-1}.
\end{equation}
For the standardized feature vector $\widetilde{\boldsymbol\phi}_{*,\gamma}$ of a new compound, the observation-predictive moments are
\begin{align}
\mu_{*,\gamma}&=\bar y_{\rm train}+\widetilde{\boldsymbol\phi}_{*,\gamma}^{T}\mathbf m_\gamma,\\
v_{*,\gamma}&=\sigma^2+\widetilde{\boldsymbol\phi}_{*,\gamma}^{T}\mathbf S_\gamma\widetilde{\boldsymbol\phi}_{*,\gamma}.
\end{align}
Here $\mathbf S_\gamma$ is the coefficient posterior covariance, and $\sigma^2$ is the observation-noise contribution. Model averaging gives
\begin{align}
\mu_{*,\rm BMA}&=\sum_\gamma w_\gamma\mu_{*,\gamma},\\
v_{*,\rm BMA}&=\sum_\gamma w_\gamma v_{*,\gamma}
+\sum_\gamma w_\gamma(\mu_{*,\gamma}-\mu_{*,\rm BMA})^2.
\end{align}
The second term measures between-support prediction dispersion. The approximate 95\% interval is $\mu_{*,\rm BMA}\pm1.96\sqrt{v_{*,\rm BMA}}$. Matching two moments does not recover the tail quantiles of a skewed or multimodal Gaussian mixture.

\subsection{Support prior and scope of uncertainty}
All admissible supports at fixed $d$ have equal prior probability,
\begin{equation}
p(\gamma)=\binom{K}{d}^{-1},\qquad |\gamma|=d.
\end{equation}
This is uniform over retained expressions, not over classes of equivalent or nearly equivalent representations. A family containing many similar expressions can receive more aggregate probability than a sparsely represented alternative. Moreover, standardization removes arbitrary column scales from the isotropic coefficient prior but does not remove dependence of the marginal likelihood on within-support correlations. Posterior weights and PIPs are therefore conditional measures of support competition under this representation and prior, rather than probabilities of microscopic mechanisms.

The plug-in noise variance is an unadjusted training residual mean square after descriptor selection. Selection from a large library can make this estimate optimistic; using training data only prevents validation leakage but does not remove selection bias. The target mean is treated as fixed, so intercept uncertainty is omitted. For comparison, integrating an intercept with a flat prior under the corresponding centered Gaussian model would add $\sigma^2/n_{\rm train}$ to the predictive variance. No such correction is applied to the reported results. Uncertainty in the estimated variance scales and in the response-dependent screening stage is also excluded. These omissions and the normal interval approximation can affect coverage; the present results do not isolate their individual contributions.

\subsection{Validation protocol}
Evaluation uses one five-fold partition with random seed 42. In every fold, standard SISSO constructs its target- and residual-driven feature union using training compounds only. Expressions are evaluated on held-out primary features without their target values. Centering, scaling, coefficient fitting, and both variance scales use the training subset. The three predictors are deterministic SISSO, the coefficient-posterior mean within the maximum-a-posteriori support, and the BMA mean. Descriptor dimensions $d=2$ and $d=3$ are evaluated separately; no posterior over dimension is constructed. Pooled out-of-fold RMSE and empirical coverage describe this partition and do not estimate variability across alternative partitions. Detailed fold results and computational timings are given in Supporting Information.
\section{Heusler-Alloy Dataset and SISSO Configuration}
\label{sec:dataset_heusler}

The dataset used for the SISSO analysis consists of full-Heusler alloys with nominal composition $\mathrm{X}_2\mathrm{YZ}$, where X is Co, Fe, or Ni. The compound list and associated materials information were taken from the University of Alabama Heusler Alloy Database~\cite{ua_heusler_db}. The dataset contains 88 compounds: 38 Co-based, 32 Fe-based, and 18 Ni-based alloys. For these $\mathrm{X}_2\mathrm{YZ}$ compounds, the two X atoms occupy equivalent positions, while Y and Z represent the other distinct sublattices.

For each of the X, Y, and Z sites, the descriptor set includes the first ionization energy ($I_E$), electronegativity ($X_P$), covalent radius ($R_c$), number of valence electrons ($N$), atomic volume ($V$), and number of $d$ electrons ($N_d$). This yields 17 site-resolved primary features after omitting the identically zero Z-site $d$-electron count. These 17 quantities are the components of the primary vector $\mathbf x$ in Section~\ref{sec:methods}; each $\Phi_j(\mathbf x)$ is a generated symbolic expression of those components, not an additional primary input.

\begin{figure}[t]
\centering
\includegraphics[width=0.70\textwidth]{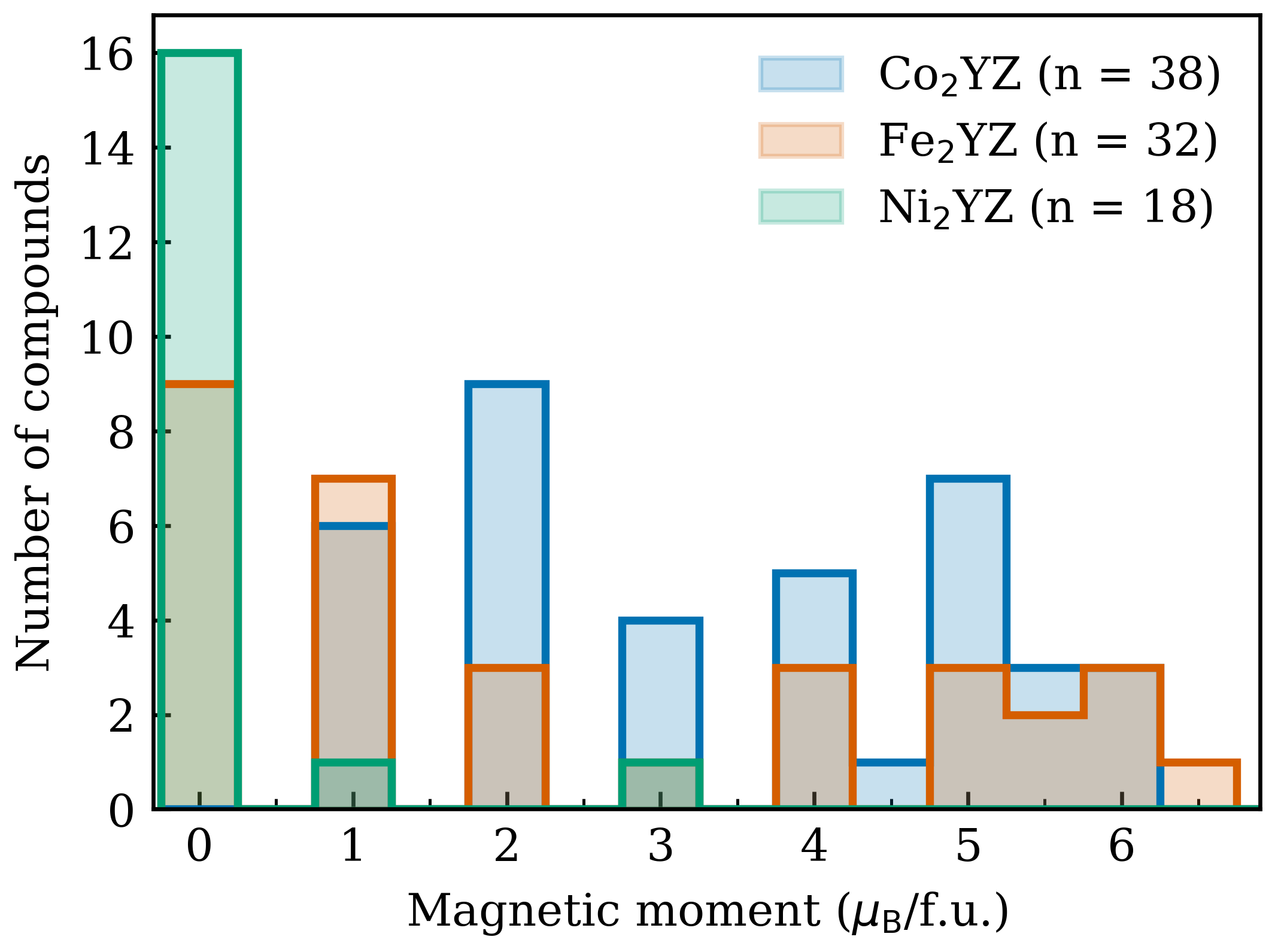}
\caption{Magnetic-moment distribution of the 88 full-Heusler alloys, separated by X-site family: 38 Co-based, 32 Fe-based, and 18 Ni-based compounds. The Ni subset is concentrated near zero moment, whereas Co and Fe subsets span broader ranges.}
\label{fig:dataset_summary}
\end{figure}

Figure~\ref{fig:dataset_summary} summarizes the distribution of magnetic moments. The histogram shows a clear separation among the three X-site families. Co-based alloys typically span 1 to 6 $\mu_\mathrm{B}$/f.u. (median 3.58 $\mu_\mathrm{B}$/f.u.). Fe-based alloys show a broader distribution, including both nearly non-magnetic and high-moment compounds (up to 6.60 $\mu_\mathrm{B}$/f.u.). The Ni-based subset is dominated by zero-moment compounds.

The primary-feature correlation heatmap (Supporting Information, Figure S1) reveals substantial multicollinearity among the site-resolved elemental descriptors. Clear blocks of positive and negative correlation appear among chemically related variables across the X, Y, and Z sublattices, especially for descriptors associated with valence, atomic size, and $d$-electron character. This inter-feature correlation is important for symbolic regression because many candidate expressions are built from inputs that are not statistically independent.

The largest target correlations occur among Y-site descriptors. Such correlations are compatible with a role for transition-metal electronic structure in Heusler magnetism,\cite{Kubler1983Formation,galanakis2002slater} but do not identify site-resolved magnetic contributions or a causal mechanism in this pooled dataset.
The strongest positive correlations are $N_Y$ (0.873), $I_{E,Y}$ (0.834), and $X_{P,Y}$ (0.629), whereas the strongest negative correlations are $N_{d,Y}$ (-0.859), $R_{c,Y}$ (-0.845), and $V_Y$ (-0.659). The strong negative correlation of $N_{d,Y}$ with magnetic moment may therefore reflect how progressive filling of the Y-site d shell changes the X-Y hybridization pattern~\cite{galanakis2014voids} and the occupation of spin-resolved d-derived states, rather than a simple one-variable causal effect.
In contrast, the reported X- and Z-site descriptors are noticeably weaker, with the largest magnitude among them coming from $I_{E,X}$ (0.490). Taken together, the heatmap and target-correlation ranking indicate that the Y sublattice carries the most direct correlation for the magnetic moment, while the full descriptor set remains strongly correlated internally.

The concentration of Ni-based compounds near zero moment differs from the broadly distributed magnetic targets in the Co- and Fe-based families. A zero total moment alone does not distinguish a nonmagnetic state from compensated magnetic order. A pooled continuous linear model with common Gaussian noise is used to examine support ambiguity across this chemically heterogeneous set, rather than to resolve these magnetic regimes. It does not impose nonnegative predictions or account explicitly for the concentration at zero; subgroup residuals therefore provide an important check of its limitations.

\subsection{Standard SISSO setup}

The standard SISSO calculations were carried out using the same primary
feature set for both descriptor dimensions. The regression type was set to
single-task regression. The full-data analysis uses $n_{\rm s}=88$ samples; validation calculations use only the training subset in each fold. The symbolic feature space was generated from
the primary features using the operators
\[
(+),\;(-),\;(*),\;(/),\;(\cdot)^{-1},\;(\cdot)^2,
\]
with reported rung depth three and a reported maximum feature complexity of five. Rung depth describes recursive operator application; a complexity cutoff is a separate implementation constraint. The feature-value range was restricted by 1.0e-3 and 1.0e5. SISSO unit groups constrain addition and subtraction within generated coordinates. Different coordinates can have different dimensions, which are absorbed by their fitted coefficients; a common physical unit is not required across all coordinates.

The sparsifying-operator stage used \(\ell_0\) optimization,
 with an intercept included and RMSE as the selection metric.

The SIS screening depth was specified per descriptor dimension. For the
two-dimensional search, 200 ranked features were retained in each SIS step, and
the union of the two screened sets produced a Bayesian support space with
\(K=400\) candidate features. For the three-dimensional search, 100 ranked
features were retained in each SIS step, giving \(K=300\) candidate features in
the union of the three screened sets. We therefore distinguish the
per-step SIS retention depth from \(K\), the total screened feature count used
for Bayesian enumeration. The resulting support spaces,
\(\binom{400}{2}=7.98\times10^4\) and
\(\binom{300}{3}=4.4551\times10^6\), retain the full deterministic SISSO
screened unions from the corresponding runs while remaining tractable for exact
posterior enumeration.

\section{Results and Discussion}\subsection{Full-data support competition}

The full-data symbolic library contains $2.425\times10^8$ candidate expressions before screening. The retained unions contain $K=400$ features at $d=2$ and $K=300$ at $d=3$, corresponding to 79,800 and 4,455,100 supports. Full-data fits characterize conditional support competition; held-out predictions are evaluated separately below. All magnetic-moment errors are reported in $\momentunit$.

On the full training set, standard SISSO selects the two-dimensional support \((8,203)\) with RMSE \(0.5150\), and the three-dimensional support \((8,101,203)\) with RMSE \(0.4115\). The two-dimensional feature labeled 8,
\begin{equation}
\phi_8=((V_X)^2N_X)/(N_X+N_Y+N_Z),
\end{equation}
combines X-site volume and valence information. Its denominator is the unweighted site sum $N_X+N_Y+N_Z$, not the stoichiometric full-Heusler count $N_v=2N_X+N_Y+N_Z$. It therefore cannot be identified with the Slater--Pauling electron count. The coefficient multiplying the unstandardized coordinate has units of magnetic moment divided by volume squared when electron counts are treated as dimensionless.

Table~\ref{tab:full_posterior} summarizes the standardized uniform-prior full-data
posterior. The MAP support happens to equal the deterministic support at both
dimensions under the specified standardized model, but the posterior is not concentrated on a
unique model: eight supports are required to reach 95\% probability in each
case. At $d=2$, the leading PIPs are 0.883, 0.559, and 0.215 for features 8,
203, and 201 in the two-dimensional union; at $d=3$, the leading values are 0.998, 0.836, 0.427, 0.289,
and 0.237 for features 8, 101, 203, 202, and 201. The agreement of the two
point supports on the full dataset therefore does not imply descriptor
uniqueness or a general equivalence between MAP and deterministic SO.

\begin{table}[t]
\centering
\caption{Uniform-prior full-data posterior after symbolic-feature
standardization. $N_{95}$ is the number of leading supports required to reach
95\% posterior probability; $O_{12}$ is the posterior odds of the first support
relative to the second.}
\label{tab:full_posterior}
\begin{tabular}{@{}cccccc@{}}
\toprule
$d$ & Deterministic & MAP & $p_{\rm MAP}$ & $O_{12}$ & $N_{95}$ \\
\midrule
2 & $(8,203)$ & $(8,203)$ & 0.526 & 2.48 & 8 \\
3 & $(8,101,203)$ & $(8,101,203)$ & 0.366 & 1.53 & 8 \\
\bottomrule
\end{tabular}
\end{table}

Integer labels refer to the corresponding run-specific feature union. In particular, equal labels in the two- and three-dimensional calculations do not establish equality of expressions. Only the two-dimensional formulas are interpreted algebraically here. High inclusion probability identifies a recurring coordinate within a specified screen and prior; it does not establish a microscopic law.

\begin{figure*}[t]
\centering
\includegraphics[width=\textwidth,height=0.42\textheight,keepaspectratio]{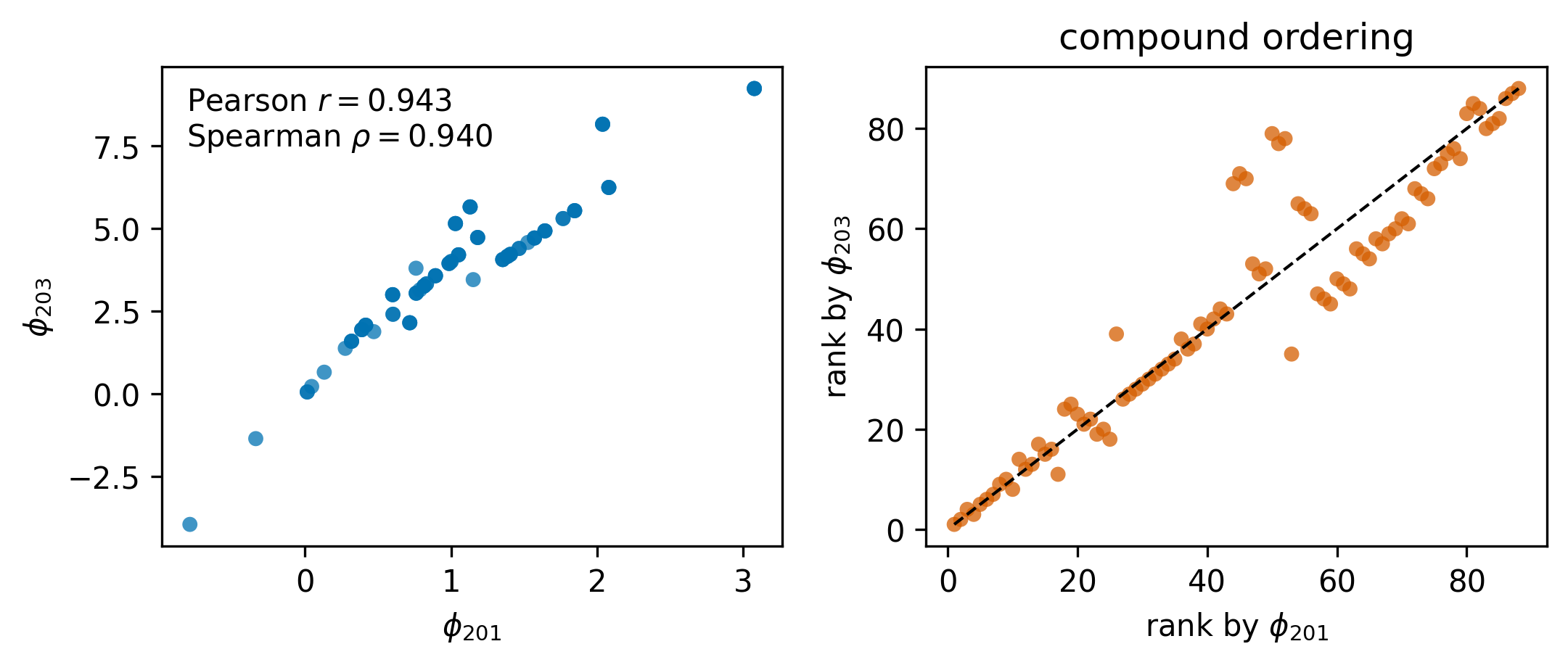}
\caption{Numerical comparison of full-data two-dimensional SISSO features 201
and 203. Left: raw feature values over 88 compounds. Right: compound ranks
induced by each feature; the dashed line denotes identical ordering. The high
but nonunit Pearson and Spearman correlations quantify functional similarity
without implying physical equivalence.}
\label{fig:feature_201_203}
\end{figure*}

The second coordinate in the \(d=2\) descriptor belongs to the 201/203 family,
\begin{equation}
\phi_{201}=((N_XV_Y)/(N_ZV_X))-(N_{d,Y}/N_{d,X}),
\label{des-201}
\end{equation}
and
\begin{equation}
\phi_{203}=((N_XV_Y)/V_X)-((N_ZN_{d,Y})/N_{d,X}).
\label{des-203}
\end{equation}
Across the 88 compounds, these coordinates are strongly but not perfectly
associated: their Pearson correlation is $r=0.943$ and their Spearman rank
correlation is $\rho=0.940$ [Figure~\ref{fig:feature_201_203}]. Their median
absolute rank displacement is three compounds, although the maximum is 29.
These expressions satisfy $\phi_{203}=N_Z\phi_{201}$. Because $N_Z$ varies across compounds, this is not a single constant rescaling over the dataset. The expressions often organize the data similarly and can act as
statistical substitutes, but they are neither algebraically nor functionally
identical. The analysis supports descriptor competition within a
valence/volume-contrast family; it does not establish physical equivalence.

\subsection{Fold-local predictive validation}

\begin{figure*}[t]
\centering
\includegraphics[width=0.82\textwidth]{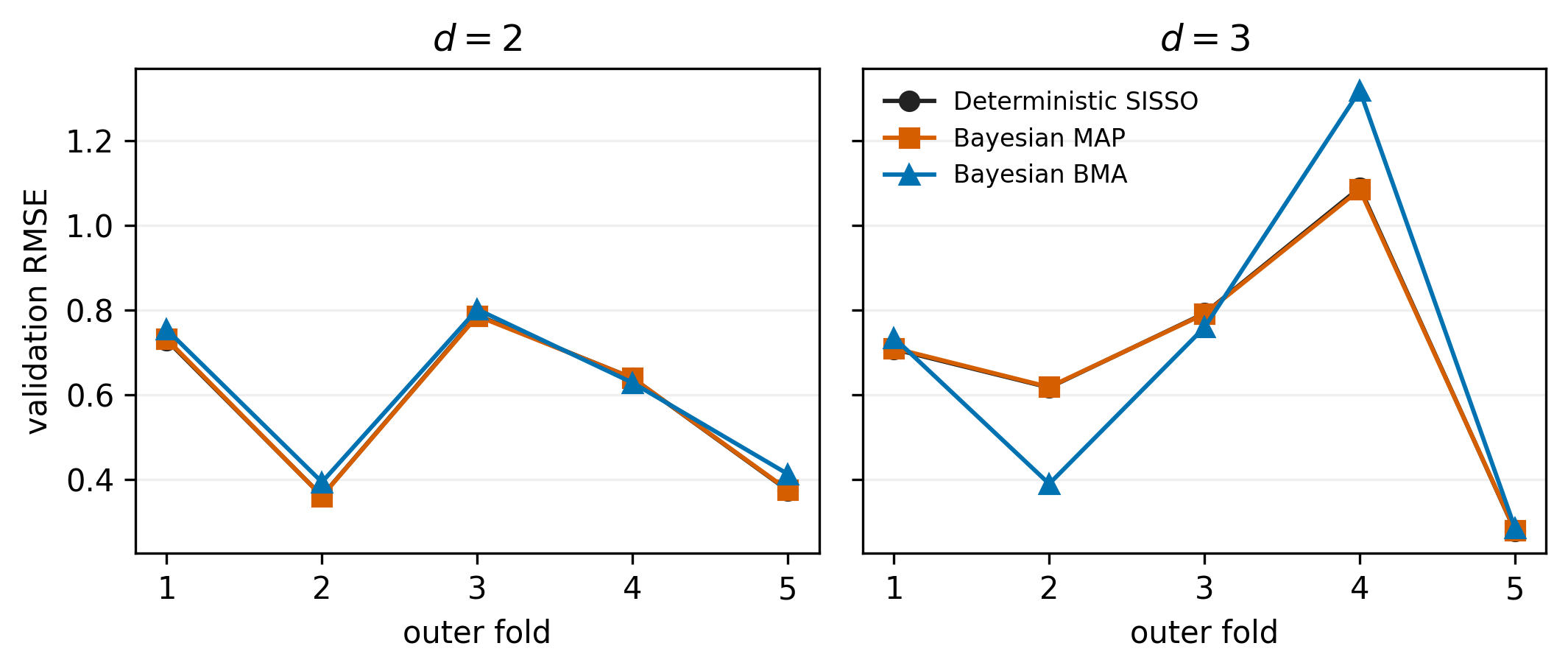}
\caption{Fold-resolved leakage-free validation RMSE (in $\momentunit$) for deterministic SISSO,
Bayesian MAP-support prediction, and Bayesian model averaging (BMA). Each
outer fold constructs its symbolic candidate union and estimates all scaling
and variance quantities using training data only.}
\label{fig:cv_rmse}
\end{figure*}

Table~\ref{tab:cv_comparison} and Figure~\ref{fig:cv_rmse} report the pooled and fold-resolved out-of-fold predictions from
the seed-42 nested five-fold calculation.  Each row is based on a fresh SISSO
run in every outer training fold; none reuses the symbolic screen generated
from all 88 targets.  Deterministic SISSO, the posterior mean conditional on
the MAP support, and BMA are shown separately.  The interval column applies
only to BMA.

\begin{table}[t]
\centering
\caption{Leakage-free nested five-fold validation under the uniform support
prior. RMSEs are pooled over all 88 out-of-fold predictions. Coverage is the
empirical coverage of the nominal 95\% BMA predictive interval.}
\label{tab:cv_comparison}
\begin{tabular}{clcc}
\toprule
$d$ & Method & RMSE ($\momentunit$) & 95\% coverage \\
\midrule
2 & deterministic SISSO & 0.6063 & --- \\
2 & Bayesian MAP & 0.6066 & --- \\
2 & Bayesian BMA & 0.6229 & 0.909 \\
3 & deterministic SISSO & 0.7430 & --- \\
3 & Bayesian MAP & 0.7422 & --- \\
3 & Bayesian BMA & 0.7812 & 0.898 \\
\bottomrule
\end{tabular}
 \end{table}

For $d=2$, the pooled RMSEs are 0.6063 for deterministic SISSO, 0.6066 for
the MAP-support predictor, and 0.6229 for BMA. Thus BMA is 0.0166 higher in
RMSE on this partition, while the MAP difference is approximately 0.0003 from the displayed pooled values. The five
deterministic and MAP supports agree by exact symbolic expression, although
two folds select expressions not present in the full-data screened union.
This observation both explains why integer feature labels cannot be compared
naively across folds and confirms that a full-data screen must not be reused
inside validation. For $d=3$, the corresponding pooled RMSEs are 0.7430,
0.7422, and 0.7812; BMA is 0.0382 higher than deterministic SISSO, whereas the
MAP difference is $-0.0008$. Deterministic and MAP supports also agree by exact
expression in all five 3D folds, giving 10/10 agreement overall in this
calculation. This empirical agreement is not a mathematical equivalence of the
selection criteria. With only one fixed partition, all differences are
descriptive rather than evidence of a significant predictive advantage.

The nominal BMA interval coverage and the residual variation are evaluated
further below. Overall coverage is 0.909 at $d=2$ and 0.898 at $d=3$, below
the nominal 0.95 level. These intervals propagate coefficient and support
uncertainty under the stated Gaussian common-noise model; their empirical
coverage in one five-fold partition should not be called calibration in the
stronger repeated-sampling sense.

Increasing the descriptor dimension reduces full-data training RMSE from 0.5150 to 0.4115, while the pooled deterministic validation RMSE rises from 0.6063 to 0.7430 and the BMA RMSE rises from 0.6229 to 0.7812. This opposing trend is consistent with overfitting at $d=3$ on the evaluated partition. Fold 4 is especially sensitive: deterministic and BMA RMSEs are 1.0886 and 1.3184 at $d=3$, compared with 0.6387 and 0.6278 at $d=2$. Averaging supports of a fixed cardinality does not select a smaller descriptor dimension, and the present BMA calculation does not remove this degradation. A comparison across dimensions would require an explicit dimension prior and mutually comparable likelihood and variance assumptions; separately normalized fixed-$d$ posteriors do not provide that comparison.

\subsection{Prescribed-candidate synthetic check}

\begin{figure*}[t]
\centering
\includegraphics[width=0.88\textwidth]{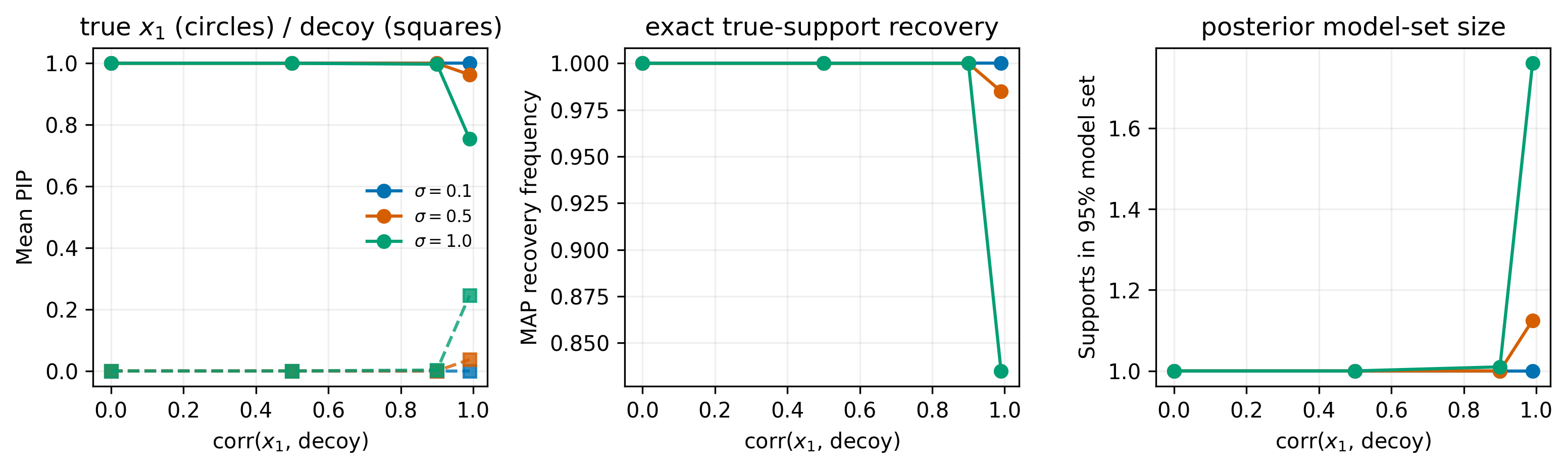}
\caption{Prescribed-candidate synthetic check over 200 realizations per condition, using the known generating noise variance and no SISSO screening. The
decoy feature is constructed with controlled correlation to true feature
$x_1$. Left: mean PIPs of $x_1$ (circles) and the decoy (squares). Center: MAP
recovery frequency of the exact true support $(x_1,x_2)$. Right: mean number of
supports required to reach 95\% posterior probability. Probability sharing at
high collinearity is the expected uncertainty-aware result.}
\label{fig:synthetic_recovery}
\end{figure*}

The fixed-screen inference stage is examined using the controlled model,
$y=1.5x_1-x_2+\varepsilon$, with candidate features $x_1$, $x_2$, an
independent nuisance feature, and a decoy whose correlation with $x_1$ was
varied. Each noise--correlation condition contains 200 independent realizations
of 80 samples. There are four prescribed candidate columns and six two-feature supports. The generating noise variance is supplied exactly, and no symbolic expansion or SISSO screening is performed. This is a check of conditional support inference under collinearity, not an end-to-end validation of screened symbolic regression or of plug-in noise estimation. The full protocol and condition-wise results are given in Supporting Information, Section S4. At correlations up to 0.9 the MAP support recovers the two true
features in essentially every realization over the tested noise levels. At
correlation 0.99 and noise standard deviation 1.0, posterior inclusion is shared
between $x_1$ and its decoy (mean PIPs 0.755 and 0.245), and exact MAP recovery
falls to 0.835. Nevertheless, the true support lies in the 95\% posterior model
set in 0.995 of realizations. Figure~\ref{fig:synthetic_recovery} demonstrates
the intended behavior: uncertainty is concentrated on statistically
indistinguishable alternatives rather than hidden by a single point support.

\subsection{Conditional feature inclusion}

\begin{figure*}[t]
\centering
\includegraphics[width=0.80\textwidth]{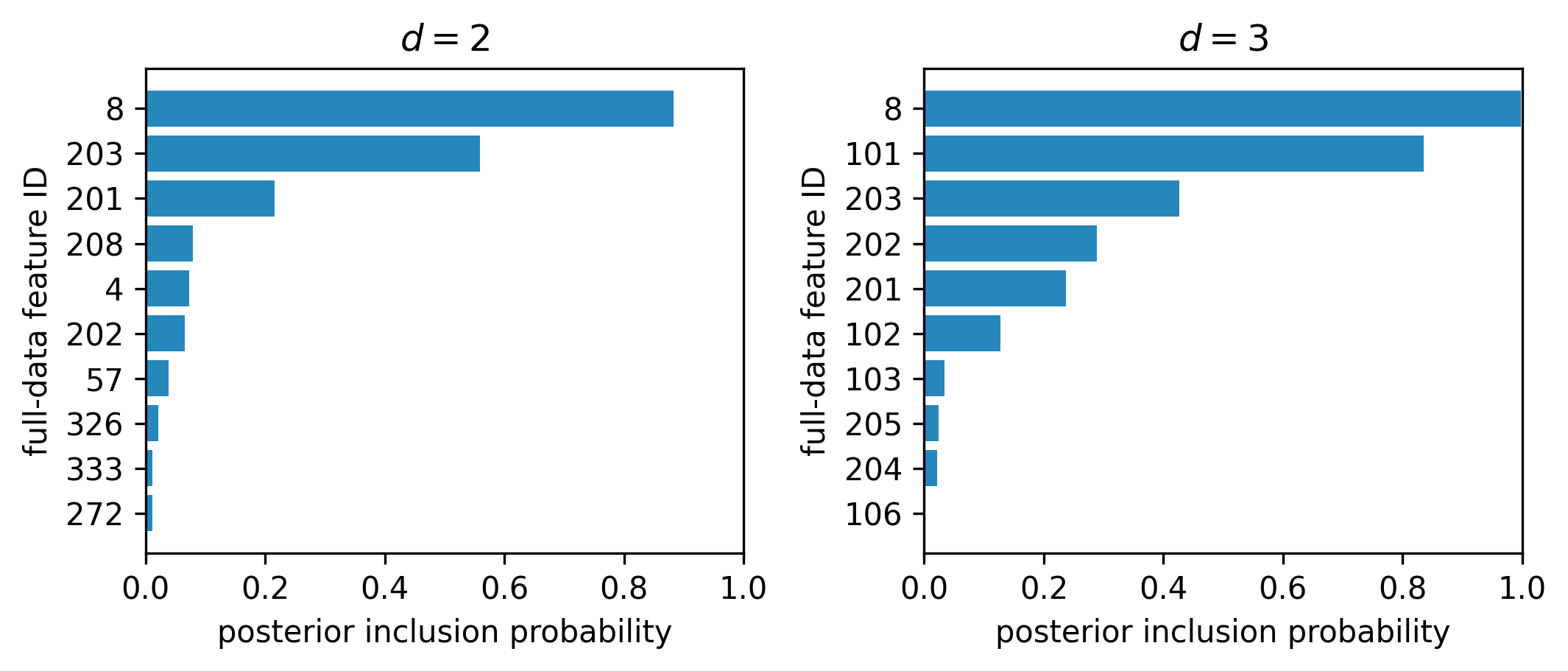}
\caption{Ten largest full-data posterior inclusion probabilities (PIPs) under
the uniform support prior. Feature identifiers are local to the corresponding
full-data SISSO union; comparison across dimensions requires comparison of the
symbolic expressions rather than the integer labels.}
\label{fig:full_data_pips}
\end{figure*}

Figure~\ref{fig:full_data_pips} shows the leading full-data PIPs under the
primary uniform prior. The posterior mass is concentrated in a small family,
but not on one algebraically unique descriptor. In particular, the two-dimensional features 201 and 203 retain competing inclusion probability. Several labels also share appreciable probability in the three-dimensional analysis. The $d=3$ feature labels denote that run's union and are not assigned the two-dimensional formulas solely on the basis of equal integers.

\subsection{Residual and subgroup diagnostics}

\begin{figure*}[t]
\centering
\includegraphics[width=0.80\textwidth]{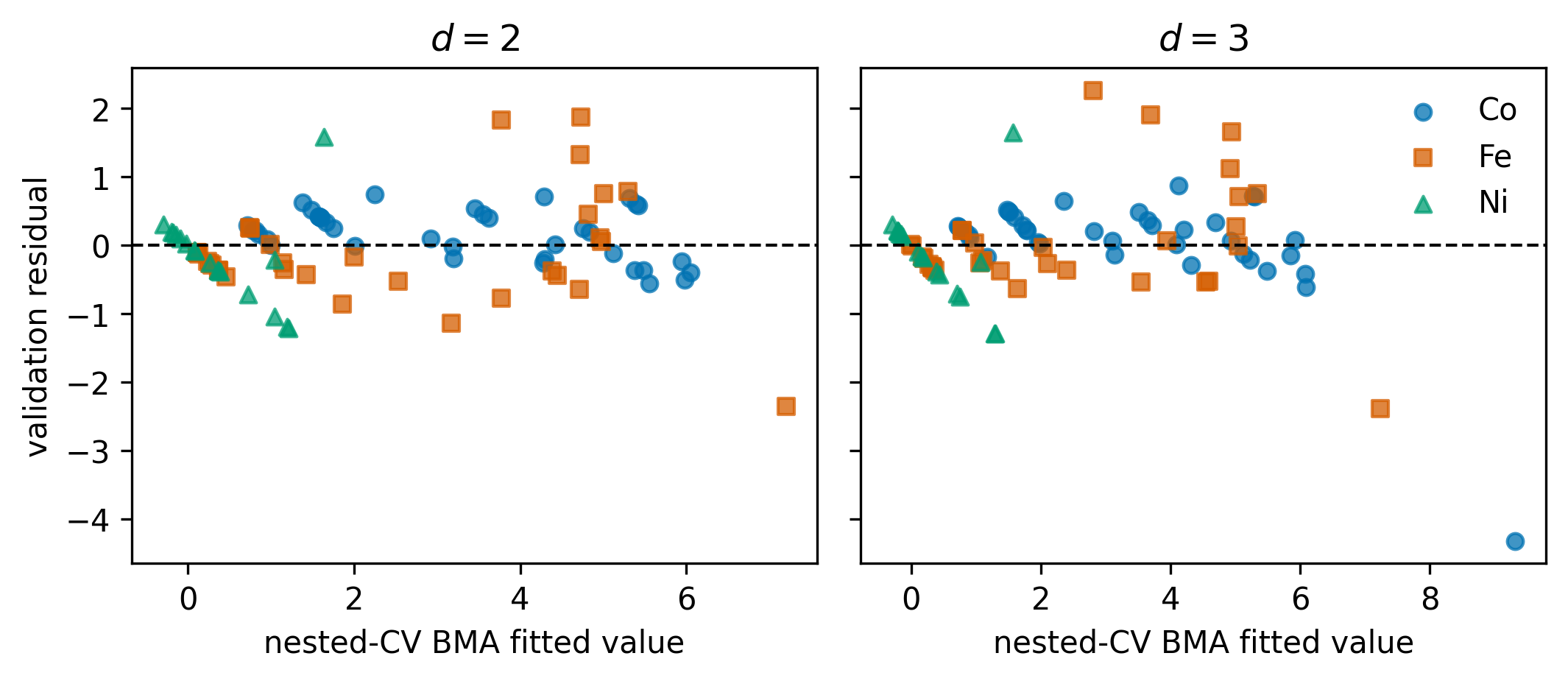}
\caption{Out-of-fold Bayesian-model-averaged (BMA) residuals versus fitted
values for the seed-42 nested five-fold calculation, separated by X-site
family. Fitted values and residuals are in $\momentunit$.}
\label{fig:subgroup_residuals}
\end{figure*}

Figure~\ref{fig:subgroup_residuals} compares nested-CV BMA residuals for the
Co-, Fe-, and Ni-based families. At $d=2$, residual standard deviations are
0.363, 0.807, and 0.647, with coverages 1.000, 0.844, and 0.833,
respectively. At $d=3$, the corresponding values are 0.800, 0.842, and 0.666,
with coverages 0.974, 0.844, and 0.833. The family dependence and overall
undercoverage make the common-noise model questionable, especially for the
Fe- and Ni-based subsets. A Co-based $d=3$ prediction above $8\,\momentunit$ has a residual below $-4\,\momentunit$, illustrating the sensitivity of a sparse fit that predicts beyond the observed full-dataset target range. Common-noise misspecification, optimistic plug-in noise, omitted intercept and variance-scale uncertainty, and the normal mixture approximation are plausible contributors to undercoverage; these results do not determine their relative importance. Family-dependent or hurdle models would change the statistical model and require separate evaluation.

\section{Conclusion}
Conditional Bayesian enumeration reveals that a deterministic SISSO descriptor can coexist with several statistically competitive alternatives. For the 88-compound Heusler dataset, the most probable support matches the deterministic selection at both examined dimensions, yet eight supports are needed to reach 95\% posterior probability. The explicit two-dimensional formulas illustrate how correlated valence and volume combinations can distribute probability across distinct symbolic representations. This ambiguity qualifies the interpretation of one selected expression as a uniquely identified physical relationship.

The predictive results delimit this contribution. On the single evaluated five-fold partition, BMA increases pooled RMSE and its approximate nominal 95\% intervals under-cover. Increasing descriptor dimension improves the training fit while degrading held-out predictions. Thus conditional support averaging provides a diagnostic of representation ambiguity, not demonstrated predictive improvement or calibrated end-to-end uncertainty. Screening variability, prior sensitivity, repeated-partition evaluation, and a noise model compatible with distinct magnetic regimes remain necessary for broader uncertainty claims.
\section*{Supporting Information}
Supporting Information contains primary-feature correlations, leading full-data support probabilities, fold-resolved prediction errors, computational timings, and the prescribed-candidate synthetic protocol.

\section*{Acknowledgements}
This work was supported by KIST (Korea Institute of Science and Technology) Institutional Program: Global Knowledge Platform Research Project (26Z0100).

\section*{Conflict of Interest}
The author declares no conflict of interest. 

\section*{Data and Software Availability}
The Heusler alloy data, SISSO results, and the codes  for performing the Bayesian analysis are freely available at:\\
https://github.com/sata-deep/Uncertainty-Aware-Symbolic-Regression-using-SISSO-supports

\bibliography{references}

\begin{tocentry}
\includegraphics[width=9cm]{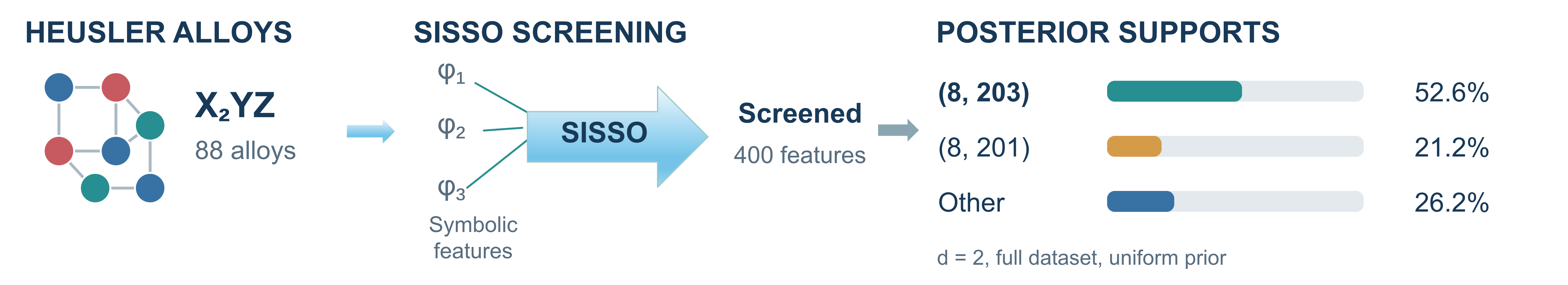}

A compact materials descriptor need not be uniquely determined by the data. Conditional Bayesian analysis of screened symbolic expressions reveals competing representations of Heusler magnetic moments.
\end{tocentry}
\clearpage
\input{supporting_information.tex}

\end{document}

%% file: supporting_information.tex
\begin{center}\Large\bfseries Supporting Information\par\vspace{1em}Descriptor Non-Uniqueness in Symbolic Regression: Conditional Bayesian Analysis of SISSO Supports\par\vspace{1em}\normalsize Satadeep Bhattacharjee\end{center}
\section{Computational Cost and Primary-Feature Correlations}
The enumeration cost grows as $\binom{K}{d}$. In the reported five-fold calculation, fold-local SISSO construction takes 2131.5 s at $d=2$ and 3252.3 s at $d=3$; the corresponding uniform-prior Bayesian calculations take 1.0 and 48.2 s. The combined calculation therefore takes approximately 1.50 h, dominated by symbolic feature construction. These are run-specific timings; hardware and parallelization details are not specified, so they do not define portable performance benchmarks. Exact enumeration is practical for the reported $K$ and $d$, but its combinatorial cost limits extrapolation to larger support spaces.

\begin{figure}[p]
\centering
\includegraphics[width=0.95\textwidth]{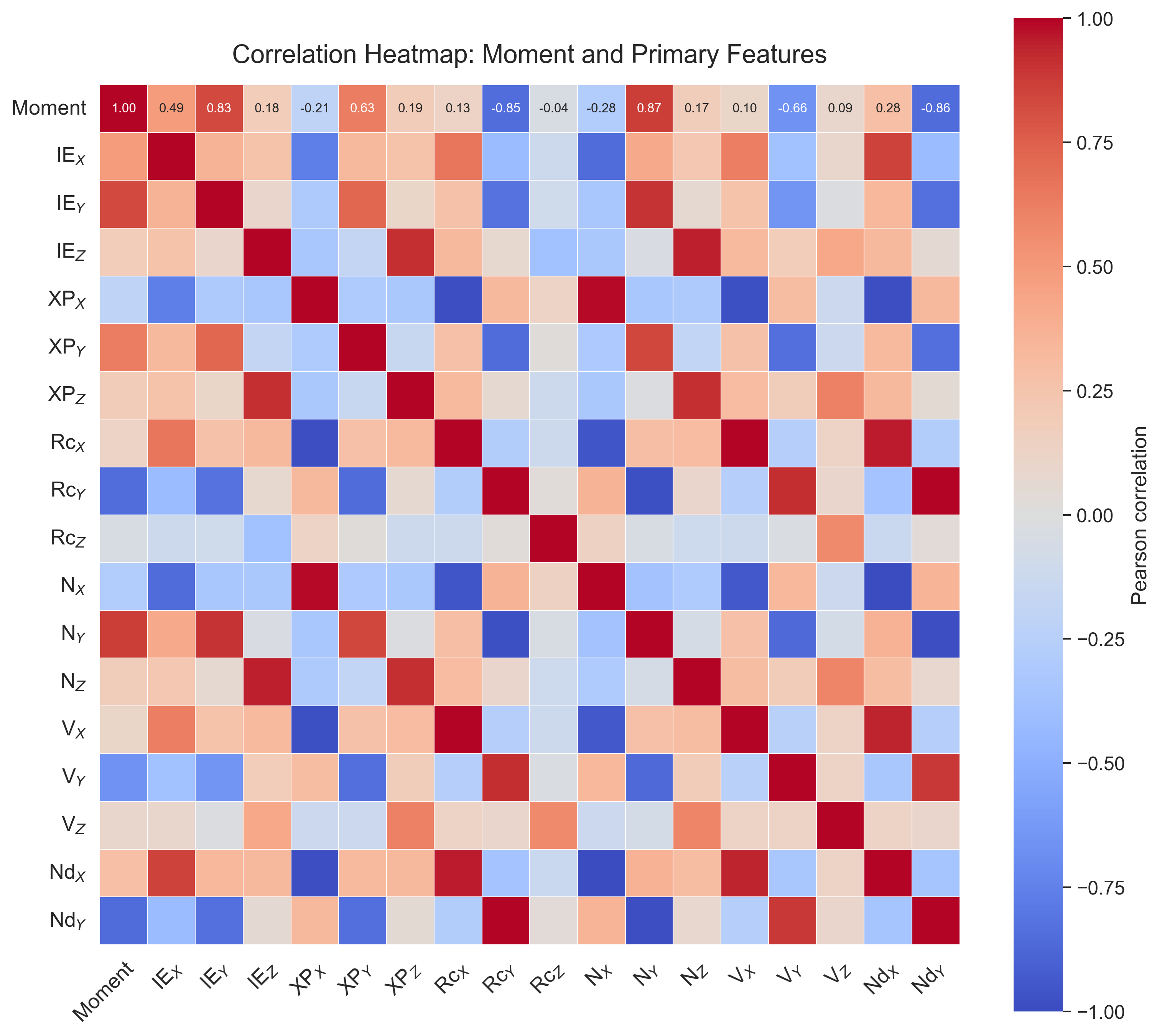}
\caption{Primary-feature and target Pearson-correlation heatmap. Labels using $Z'$ refer to the Z-site quantities denoted by $Z$ in the manuscript. Correlations summarize this dataset and do not establish causal or site-resolved magnetic contributions.}
\end{figure}
\section{Leading full-data support probabilities}

Table~\ref{tab:top_supports_appendix} gives the ten leading supports from the
standardized uniform-prior posterior. Feature IDs are local to the corresponding
full-data SISSO union and should not be compared across descriptor dimensions
without their expressions.

\begin{table}[H]
\centering
\caption{Leading uniform-prior supports and cumulative posterior probability
for the standardized full-data analyses.}
\label{tab:top_supports_appendix}
\begin{minipage}{\textwidth}
\centering
\textbf{$d=2$}\\[0.4em]
\begin{tabular}{rllr}
\toprule
Rank & Support & $p(\gamma\mid\mathcal D)$ & Cumulative \\
\midrule
1 & $(8,203)$ & 0.5264 & 0.5264 \\
2 & $(8,201)$ & 0.2119 & 0.7383 \\
3 & $(8,208)$ & 0.07386 & 0.8122 \\
4 & $(8,202)$ & 0.06418 & 0.8764 \\
5 & $(57,203)$ & 0.0314 & 0.9078 \\
6 & $(4,326)$ & 0.02132 & 0.9291 \\
7 & $(4,333)$ & 0.01174 & 0.9408 \\
8 & $(4,272)$ & 0.01115 & 0.952 \\
9 & $(4,339)$ & 0.006205 & 0.9582 \\
10 & $(4,282)$ & 0.006059 & 0.9643 \\
\bottomrule
\end{tabular}
 \end{minipage}\par\vspace{1.5em}
\begin{minipage}{\textwidth}
\centering
\textbf{$d=3$}\\[0.4em]
\begin{tabular}{rllr}
\toprule
Rank & Support & $p(\gamma\mid\mathcal D)$ & Cumulative \\
\midrule
1 & $(8,101,203)$ & 0.3660 & 0.3660 \\
2 & $(8,101,202)$ & 0.2399 & 0.6059 \\
3 & $(8,101,201)$ & 0.1895 & 0.7954 \\
4 & $(8,102,203)$ & 0.05419 & 0.8496 \\
5 & $(8,102,202)$ & 0.03561 & 0.8852 \\
6 & $(8,102,201)$ & 0.03102 & 0.9162 \\
7 & $(8,101,205)$ & 0.02131 & 0.9375 \\
8 & $(8,101,204)$ & 0.01811 & 0.9556 \\
9 & $(8,103,201)$ & 0.01502 & 0.9706 \\
10 & $(8,103,202)$ & 0.01167 & 0.9823 \\
\bottomrule
\end{tabular}
 \end{minipage}
\end{table}

\section{Fold-resolved nested validation}

Table~\ref{tab:fold_supports_appendix} reports each outer-fold error and coverage. Deterministic and MAP supports agree by exact symbolic expression in all ten dimension--fold combinations, in the reported calculation. In two folds at $d=2$ and four folds at $d=3$, selected expressions are absent from the corresponding full-data union. Feature integers are run-specific identifiers and are not interchangeable across folds.

\begin{table}[H]
\centering
\small
\caption{Fold-resolved deterministic, MAP-support, and BMA results from the
fold-local uniform-prior calculation. RMSE is in $\momentunit$. Coverage is for nominal 95\%
BMA predictive intervals.}
\label{tab:fold_supports_appendix}
\begin{tabular}{cccccc}
\toprule
$d$ & Fold & RMSE$_{\rm det}$ & RMSE$_{\rm MAP}$ & RMSE$_{\rm BMA}$ & Coverage \\
\midrule
2 & 1 & 0.7286 & 0.7310 & 0.7535 & 0.778 \\
2 & 2 & 0.3603 & 0.3599 & 0.3931 & 1.000 \\
2 & 3 & 0.7870 & 0.7855 & 0.8013 & 0.833 \\
2 & 4 & 0.6387 & 0.6391 & 0.6278 & 0.941 \\
2 & 5 & 0.3733 & 0.3746 & 0.4120 & 1.000 \\
3 & 1 & 0.7072 & 0.7095 & 0.7342 & 0.722 \\
3 & 2 & 0.6166 & 0.6177 & 0.3889 & 1.000 \\
3 & 3 & 0.7918 & 0.7905 & 0.7588 & 0.889 \\
3 & 4 & 1.0886 & 1.0842 & 1.3184 & 0.882 \\
3 & 5 & 0.2769 & 0.2783 & 0.2837 & 1.000 \\
\bottomrule
\end{tabular}
 \end{table}

For each fold, define
$\Delta_f=\operatorname{RMSE}_{\rm BMA,f}-\operatorname{RMSE}_{\rm det,f}$,
so positive values favor deterministic SISSO. Table~\ref{tab:paired_differences}
uses the unrounded machine-readable RMSEs. The reported
$s_\Delta/\sqrt{5}$ is only a descriptive dispersion measure: the five
training sets overlap, so it is not a validated standard error of
generalization performance. The mean fold-wise difference also differs from
the difference between pooled RMSEs because the latter pools squared residuals
before taking the square root.

\begin{table}[H]
\centering
\caption{Paired fold differences $\Delta_f$ in
$\mu_{\rm B}$/f.u., calculated from the unrounded nested-CV outputs.}
\label{tab:paired_differences}
\begin{tabular}{crr}
\toprule
Fold & $d=2$ & $d=3$ \\
\midrule
1 & 0.02491 & 0.02695 \\
2 & 0.03281 & $-0.22761$ \\
3 & 0.01426 & $-0.03306$ \\
4 & $-0.01096$ & 0.22979 \\
5 & 0.03869 & 0.00676 \\
\midrule
Mean & 0.01994 & 0.00056 \\
$s_\Delta/\sqrt{5}$ & 0.00874 & 0.07296 \\
\bottomrule
\end{tabular}
 \end{table}

\section{Known-support synthetic protocol}
\label{app:synthetic_protocol}

For every pair $\sigma_\varepsilon\in\{0.1,0.5,1.0\}$ and
$\rho\in\{0,0.5,0.9,0.99\}$, 200 data sets of $n_{\rm s}=80$ were generated
from independent standard-normal $x_1$, $x_2$, and nuisance variables. The
decoy was $x_{\rm decoy}=\rho x_1+\sqrt{1-\rho^2}\,z$, with independent
standard-normal $z$, and the response was
$y=1.5x_1-x_2+\varepsilon$, $\varepsilon\sim\mathcal N(0,\sigma_\varepsilon^2)$.
The base seed was 20260818; each realization used the deterministic seed
$20260818+r+10000\sigma_\varepsilon+10^6\rho$. All four prescribed columns
were standardized, the six two-feature supports had a uniform prior,
$\sigma^2=\sigma_\varepsilon^2$, and
$\tau^2=\operatorname{var}(y)/2$ using the population variance. This test did
not regenerate a symbolic library or rerun residual-driven SISSO and therefore
assesses the fixed-screen support-inference stage, not the end-to-end pipeline.

\begin{table}[H]
\caption{Condition-wise results of the prescribed-candidate synthetic test. $\rho$ is the true correlation between $x_1$ and its decoy; $N_{95}$ is the number of leading supports needed to reach 95\% posterior probability.}
\label{tab:synthetic_prescribed}
\centering
\begin{tabular}{ccccc}
\toprule
$\sigma_\varepsilon$ & $\rho$ & MAP recovery & True in 95\% set & Mean $N_{95}$ \\
\midrule
0.1 & 0.00 & 1.000 & 1.000 & 1.000 \\
0.1 & 0.50 & 1.000 & 1.000 & 1.000 \\
0.1 & 0.90 & 1.000 & 1.000 & 1.000 \\
0.1 & 0.99 & 1.000 & 1.000 & 1.000 \\
0.5 & 0.00 & 1.000 & 1.000 & 1.000 \\
0.5 & 0.50 & 1.000 & 1.000 & 1.000 \\
0.5 & 0.90 & 1.000 & 1.000 & 1.000 \\
0.5 & 0.99 & 0.985 & 0.990 & 1.125 \\
1.0 & 0.00 & 1.000 & 1.000 & 1.000 \\
1.0 & 0.50 & 1.000 & 1.000 & 1.000 \\
1.0 & 0.90 & 1.000 & 1.000 & 1.010 \\
1.0 & 0.99 & 0.835 & 0.995 & 1.760 \\
\bottomrule
\end{tabular}
\end{table}

%% file: references.bib
@article{ouyang2018sisso,
  title={SISSO: A compressed-sensing method for identifying the best low-dimensional descriptor in an immensity of offered candidates},
  author={Ouyang, Runhai and Curtarolo, Stefano and Ahmetcik, Emre and Scheffler, Matthias and Ghiringhelli, Luca M},
  journal={Physical Review Materials},
  volume={2},
  number={8},
  pages={083802},
  year={2018},
  publisher={APS}
}

@article{desmond2026exhaustive,
  title={(Exhaustive) symbolic regression and model selection by minimum description length},
  author={Desmond, Harry},
  journal={Philosophical Transactions of the Royal Society A: Mathematical, Physical and Engineering Sciences},
  volume={384},
  number={2317},
  pages={20240584},
  year={2026},
  publisher={Royal Society Publishing}
}

@article{10.1098/rsta.2025.0089,
    author = {Guimer{\`a}, Roger and Sales-Pardo, Marta},
    title = {Bayesian symbolic regression: automated equation discovery from a physicist's perspective},
    journal = {Philosophical Transactions of the Royal Society A: Mathematical, Physical and Engineering Sciences},
    volume = {384},
    number = {2317},
    pages = {20250089},
    year = {2026},
    month = {04},
    issn = {1364-503X},
    doi = {10.1098/rsta.2025.0089},
    url = {https://doi.org/10.1098/rsta.2025.0089},
    eprint = {https://royalsocietypublishing.org/rsta/article-pdf/doi/10.1098/rsta.2025.0089/6131240/rsta.2025.0089.pdf},
}

@article{Kubler1983Formation,title={Formation and coupling of magnetic moments in Heusler alloys},author={J. K{\"u}bler and A. R. Williams and C. Sommers},journal={Physical Review B},year={1983},volume={28},pages={1745-1755},doi={10.1103/physrevb.28.1745}}

@article{hoeting1999bayesian,
  title={Bayesian model averaging: a tutorial (with comments by M. Clyde, David Draper and E. I. George, and a rejoinder by the authors)},
  author={Hoeting, Jennifer A and Madigan, David and Raftery, Adrian E and Volinsky, Chris T},
  journal={Statistical science},
  volume={14},
  number={4},
  pages={382--417},
  year={1999},
  publisher={Institute of Mathematical Statistics}
}

@article{nair2025materials,
  title={Materials-discovery workflow guided by symbolic regression for identifying acid-stable oxides for electrocatalysis},
  author={Nair, Akhil S and Foppa, Lucas and Scheffler, Matthias},
  journal={npj Computational Materials},
  volume={11},
  number={1},
  pages={150},
  year={2025},
  publisher={Nature Publishing Group UK London}
  
}

@article{nair2026interpretable,
  title={Interpretable Bayesian Optimization for Catalyst Discovery},
  author={Nair, Akhil S and Foppa, Lucas and Scheffler, Matthias},
  journal={Faraday Discussions},
  doi={10.1039/D5FD00159E},
  year={2026},
  publisher={Royal Society of Chemistry}
}

@article{xie2019functional,
  title={Functional form of the superconducting critical temperature from machine learning},
  author={Xie, Stephan R and Stewart, Gregory R and Hamlin, James J and Hirschfeld, Peter J and Hennig, Richard G},
  journal={Physical Review B},
  volume={100},
  number={17},
  pages={174513},
  year={2019},
  publisher={APS}
}

@article{cao2020artificial,
  title={Artificial intelligence for high-throughput discovery of topological insulators: The example of alloyed tetradymites},
  author={Cao, Guohua and Ouyang, Runhai and Ghiringhelli, Luca M and Scheffler, Matthias and Liu, Huijun and Carbogno, Christian and Zhang, Zhenyu},
  journal={Physical Review Materials},
  volume={4},
  number={3},
  pages={034204},
  year={2020},
  publisher={APS}
}

@article{bhattacharjee2022general,
  title={A general rule for predicting the magnetic moment of Cobalt-based Heusler compounds using compressed sensing and density functional theory},
  author={Bhattacharjee, Satadeep and Lee, Seung-Cheol},
  journal={Journal of Magnetism and Magnetic Materials},
  volume={563},
  pages={169818},
  year={2022},
  publisher={Elsevier}
}

@article{ram2023combining,
  title={Combining First-Principles Modeling and Symbolic Regression for Designing Efficient Single-Atom Catalysts in the Oxygen Evolution Reaction on {Mo$_2$CO$_2$} MXenes},
  author={Ram, Swetarekha and Choi, Gwan Hyun and Lee, Albert S and Lee, Seung-Cheol and Bhattacharjee, Satadeep},
  journal={ACS Applied Materials \& Interfaces},
  volume={15},
  number={37},
  pages={43702--43711},
  year={2023},
  publisher={ACS Publications}
}

@article{george1993variable,
  title={Variable selection via Gibbs sampling},
  author={George, Edward I and McCulloch, Robert E},
  journal={Journal of the American Statistical Association},
  volume={88},
  number={423},
  pages={881--889},
  year={1993},
  publisher={Taylor \& Francis}
}

@article{george1997approaches,
  title={Approaches for Bayesian variable selection},
  author={George, Edward I and McCulloch, Robert E},
  journal={Statistica sinica},
  pages={339--373},
  year={1997},
  publisher={JSTOR}
}

@misc{ua_heusler_db,
  author = {{The University of Alabama}},
  title = {Heusler Alloy Database},
  year = {2026},
  howpublished = {Online database},
  note = {Available at \url{http://heusleralloys.mint.ua.edu/}; accessed April 29, 2026}
}

@article{galanakis2002slater,
  title={Slater-Pauling behavior and origin of the half-metallicity of the full-Heusler alloys},
  author={Galanakis, I and Dederichs, PH and Papanikolaou, N.},
  journal={Physical Review B},
  volume={66},
  number={17},
  pages={174429},
  year={2002},
  publisher={APS}
}

@article{galanakis2014voids,
  title={Voids-driven breakdown of the local-symmetry and Slater-Pauling rule in half-metallic Heusler compounds},
  author={Galanakis, I and {\c{S}}a{\c{s}}{\i}o{\u{g}}lu, E and Bl{\"u}gel, S and {\"O}zdo{\u{g}}an, Kemal},
  journal={Physical Review B},
  volume={90},
  pages={064408},
  year={2014}
}

@article{mitchell1988bayesian,
  author = {Mitchell, T. J. and Beauchamp, J. J.},
  title = {Bayesian Variable Selection in Linear Regression},
  journal = {Journal of the American Statistical Association},
  year = {1988}, volume = {83}, number = {404}, pages = {1023--1032},
  doi = {10.1080/01621459.1988.10478694}
}

@article{raftery1997bma,
  author = {Raftery, Adrian E. and Madigan, David and Hoeting, Jennifer A.},
  title = {Bayesian Model Averaging for Linear Regression Models},
  journal = {Journal of the American Statistical Association},
  year = {1997}, volume = {92}, number = {437}, pages = {179--191},
  doi = {10.1080/01621459.1997.10473615}
}

@article{barbieri2004optimal,
  author = {Barbieri, Maria Maddalena and Berger, James O.},
  title = {Optimal Predictive Model Selection},
  journal = {The Annals of Statistics},
  year = {2004}, volume = {32}, number = {3}, pages = {870--897},
  doi = {10.1214/009053604000000238}
}

@article{clyde2004model,
  author = {Clyde, Merlise and George, Edward I.},
  title = {Model Uncertainty},
  journal = {Statistical Science},
  year = {2004}, volume = {19}, number = {1}, pages = {81--94},
  doi = {10.1214/088342304000000035}
}

@article{chipman1996related,
  author = {Chipman, Hugh},
  title = {Bayesian Variable Selection with Related Predictors},
  journal = {Canadian Journal of Statistics},
  year = {1996}, volume = {24}, number = {1}, pages = {17--36},
  doi = {10.2307/3315687}
}

@article{scott2010multiplicity,
  author = {Scott, James G. and Berger, James O.},
  title = {Bayes and Empirical-Bayes Multiplicity Adjustment in the Variable-Selection Problem},
  journal = {The Annals of Statistics},
  year = {2010}, volume = {38}, number = {5}, pages = {2587--2619},
  doi = {10.1214/10-AOS792}
}

@misc{jin2019bayesianSR,
  title        = {Bayesian Symbolic Regression},
  author       = {Jin, Ying and Fu, Weilin and Kang, Jian and Guo, Jiadong and Guo, Jian},
  year         = {2019},
  eprint       = {1910.08892},
  archivePrefix= {arXiv},
  primaryClass = {stat.ME},
  url          = {https://arxiv.org/abs/1910.08892}
}

@inproceedings{pmlr-v337-boussif26a,
  title     = {Bayesian Symbolic Regression with Entropic Reinforcement Learning},
  author    = {Boussif, Oussama and Sabatier, Sylvain and Ferreira, Caio and Allauzen, Alexandre},
  booktitle = {Proceedings of the Forty-Second Conference on Uncertainty in Artificial Intelligence},
  pages     = {467--479},
  year      = {2026},
  editor    = {Chadha, Aman and Globerson, Amir and Vollmer, Sebastian},
  volume    = {337},
  series    = {Proceedings of Machine Learning Research},
  publisher = {PMLR},
  url       = {https://proceedings.mlr.press/v337/boussif26a.html}
}
